\documentstyle[bbm,amsfonts,epsfig,12pt,here]{article}
\newcommand{\smallfrac}[2] {\mbox{$\frac{#1}{#2}$}}
\newcommand {\eqref} [1] {(\ref {#1})}
\newcommand {\slsh} [1] {\not{\hbox{\kern-2pt${#1}$}}}

\newcommand{\drawsquare}[2]{\hbox{%
\rule{#2pt}{#1pt}\hskip-#2pt
\rule{#1pt}{#2pt}\hskip-#1pt
\rule[#1pt]{#1pt}{#2pt}}\rule[#1pt]{#2pt}{#2pt}\hskip-#2pt
\rule{#2pt}{#1pt}}

\newcommand{\Yfund}{\raisebox{-.5pt}{\drawsquare{6.5}{0.4}}}
\newcommand{\Ysymm}{\Yfund\hskip-0.4pt%
                    \Yfund}
\def\symm{\Ysymm}
\def\bsymm{\overline{\Ysymm}}

\def\drawbox#1#2{\hrule height#2pt
        \hbox{\vrule width#2pt height#1pt \kern#1pt
              \vrule width#2pt}
              \hrule height#2pt}

\def\Asym#1#2{\vcenter{\vbox{\drawbox{#1}{#2}
              \kern-#2pt       
              \drawbox{#1}{#2}}}}

\def\asymm{\Asym{6.4}{0.3}}

\def\basymm{\overline{\asymm}}

\newcommand {\beq} {\begin{equation}}
\newcommand {\eeq} {\end{equation}}
 \newcommand {\ber}{\begin{eqnarray*}}
 \newcommand {\eer} {\end{eqnarray*}}
\newcommand {\bea}{\begin{eqnarray}}
 \newcommand {\eea} {\end{eqnarray}}
\newcommand{\Nfour} {${\cal N}=4\ $}

\newcommand{\None}{${\cal N}=1\ $}
\newcommand{\ztwo}{${\bf Z}_2\ $}
\newcommand{\zk}{${\bf Z}_k\ $}

\def\Acknowledgements{\bigskip  \bigskip {\begin{center} \begin{large}
             \bf ACKNOWLEDGMENTS \end{large}\end{center}}}

\begin{document}
\begin{titlepage}
\begin{flushright}{CERN-TH/2003-022

TPI-MINN-03/04, UMN-TH-2129/03
}
\end{flushright}
\vskip 1cm

\centerline{{\Large \bf Exact Results in Non-Supersymmetric}}
\vskip 0.1cm
\centerline{{\Large \bf Large N Orientifold Field Theories}}
\vskip 1cm
\centerline{\large A. Armoni ${}^a$, M. Shifman ${}^{a,b}$, G. Veneziano ${}^a$}
\vskip 0.1cm
\centerline{\small adi.armoni, michael.shifman, gabriele.veneziano@cern.ch}
\vskip 0.5cm
\centerline{${}^a$ Theory Division, CERN}
\centerline{CH-1211 Geneva 23, Switzerland}
\vskip 0.5cm
\centerline{${}^b$ William I. Fine Theoretical Physics Institute, University
of Minnesota,}
\centerline{Minneapolis, MN 55455, USA$^\star$}
\vskip 1cm

\begin{abstract}
We consider non-supersymmetric large $N$ orientifold field theories.
 Specifically, we discuss a gauge theory with a Dirac
 fermion in the anti-symmetric tensor representation. We argue that, at large
$N$ and in a
large part of its bosonic sector, this theory is non-perturbatively equivalent
to \None SYM, so that exact results established in
the latter (parent) theory  also hold in the daughter orientifold theory.
 In particular, the non-supersymmetric theory has an exactly calculable
bifermion
 condensate, exactly degenerate parity doublets, and
 a vanishing cosmological constant (all this to leading order
in $1/N$).

\end{abstract}

\vspace{3cm}

\noindent
\rule{2.4in}{0.25mm}\\
$^\star$ Permanent address.
\end{titlepage}

\section{Introduction}

\noindent

 Gauge field theories at strong coupling are of great importance in
 particle physics. Exact results in gauge theories at strong coupling
have a special weight. In the last few
years supersymmetry (SUSY) proved to be a guiding principle
(see e.g. \cite{srev}) providing deep insights in gauge
dynamics. Recently it was suggested
that a large number of symmetry relations valid in supersymmetric
theories remain valid in the large $N$ limit in non-supersymmetric daughter
theories obtained from the parent one through orbifoldization
 \cite{Strassler:2001fs}. The most popular
(non-SUSY) orbifolds are {\ztwo}, or \zk in general
 \cite{Gorsky:2002wt,Tong:2002vp,Dijkgraaf:2002wr}.
In particular, in the last paper it was argued
that the
 superpotential
of SUSY gluodynamics \cite{Veneziano:1982ah},
\beq
W(S)= N (S \log S/\Lambda ^3 -S)\,,\qquad S= \lambda \lambda + ...\,,
 \label{VY}
\eeq
 can be in a sense extended to its \ztwo orbifold.
This is equivalent to the statement that the daughter
theory has $N$ vacua
labeled by a bifermion condensate
\beq
 \langle \lambda \lambda \rangle_k = N \Lambda ^3 \exp\left( i {2\pi k\over N}
\right).
 \label{condensate}
\eeq 
While the planar equivalence is certainly true
in perturbation theory \cite{Bershadsky:1998mb,Bershadsky:1998cb}
its nonperturbative status is being debated;
a number of arguments pro and con
can be found in the literature
\cite{Gorsky:2002wt,Tong:2002vp,Dijkgraaf:2002wr}.

 The purpose of this work is to present a non-supersymmetric
daughter
(different from the \zk orbifolds) for which the large $N$ equivalence
between the parent SUSY theory and the daughter non-supersymmetric one
 at  nonperturbative level rests on a more solid ground than in the case of
the \zk orbifolds.
This ``orientifold field theory''
 was suggested in Refs. \cite{Armoni:1999gc,Angelantonj:1999qg}
in a somewhat different context. One of its advantages over
more popular \zk orbifolds is the absence in it of the twisted sector.

The field content of the orientifold gauge field theory  differs from
the one of its
 parent theory, U($N$) SUSY gluodynamics, in that the gluinos are replaced by
massless Dirac fermions in the rank-two antisymmetric tensor representation of
 U($N$) (denoted by $\asymm + \basymm$). The total number of (say) left handed fermions is thus $N(N-1)$ in the
daughter theory and $N^2$ in the parent theory and agrees to leading order
in $1/N$.  Similarly,
one can discuss a theory with a Dirac fermion in the symmetric
representation ($\symm+ \bsymm$).
As we will see,
this theory lives on a brane configuration of type 0A
string theory \cite{Armoni:1999gc} which consists of NS5 branes,
D4 branes and an
orientifold plane --- hence the name ``orientifold field theory.''
The daughter theory in our case is a much closer cousin of \None SYM
than the \ztwo orbifold. Indeed, the gauge groups
in the parent and daughter theories are the same, and no rescaling
of the gauge couplings is needed.

Assuming that both
theories are in the confining regime\,\footnote{In the Higgs regime
the
$1/N$ expansion becomes more subtle, see a discussion below.},
we will show that in the large $N$ limit many results and
symmetry relations that were
obtained for \None SYM hold also for the above orientifold field
theory. One specific quantity is the bifermion condensate
\eqref{condensate}. It labels distinct vacua.
The number of vacua turns out to be the same in the parent and daughter
theories, $N$.
 Another result is the vanishing of the vacuum energy density
in the daughter theory.

We would like to emphasize that the spectrum of the
orientifold field theory {\em does
not} coincide with that of \None SYM in the large $N$ limit.
In particular, there is no SUSY. While the composite color-singlet
 hadrons of
\None are fermi-bose degenerate,    the composite color-singlet
 hadrons of
the orientifold field theory are purely bosonic.

The organization of this paper is as follows: in Sect. 2 we
present our main result ---
the perturbative and nonperturbative
equivalence.
In Sect. 3 we briefly present the string
theory realization of the theory in its conjectured relation to
M-theory. In Sect. 4 we compare our analysis and results with
those referring to \zk orbifold field theories
and summarize conclusions. 

\section{Orientifold field theory and \None SYM}

\noindent

In this section we will argue that in the $N\to\infty$
limit there is a sector in the orientifold theory
exactly identical to \None SYM and, therefore, exact
results on the IR behavior of this theory can be obtained.

The parent theory is \None SUSY gluodynamics with the gauge
group U($N$). In the large $N$ limit the U(1) factor is irrelevant.
The daughter theory has the same gauge group,
and  the same gauge coupling\,\footnote{To be more precise,
the gauge groups are almost the same.
The U(1) factor completely decouples in the parent theory,
while it does not decouple in the daughter one.
Moreover, in the former theory the \zk center of the
gauge group acts
trivially while in the latter one it is only \ztwo that acts trivially.
These distinctions are unimportant in the large $N$ limit.
}. The gluino field
$\lambda^i_j$ is replaced by two Weyl spinors
$\lambda_{[ij]}$ and $\xi^{[ij]}$. We can combine the Weyl spinors
into one Dirac spinor, either $\Psi_{[ij]}$ or $\Psi^{[ij]}$.
Note that the number of fermionic degrees of freedom
is $N^2-N$, as in the parent theory in the large $N$ limit.

The hadronic (color-singlet) sectors of the theories
are different. In the parent theory composite fermions with mass
scaling as $N^0$ exist, and moreover, they are degenerate with their
bosonic  SUSY counterparts. In the daughter theory any interpolating
color-singlet
current with the fermion quantum numbers contains a number of constituents
growing  with $N$. Hence at $N=\infty$
the spectrum contains only bosons.

Classically the parent theory has just an $R$ symmetry corresponding to
chiral rotations of the gluino field. Instantons break this symmetry down to
${\bf Z}_{2N}\ $.
The daughter theory has, on top, a conserved anomaly free  current
\beq
\bar \lambda_{\dot\alpha } \lambda_{\alpha}
-\bar \xi_{\dot\alpha } \xi_{\alpha}\,.
\label{veccur}
\eeq
In terms of the Dirac spinor this is the vector current
$\bar\Psi\gamma_\mu\Psi$.
If the corresponding charge is denoted by $Q$,
in the color-singlet bosonic sector $Q=0$,
with necessity. Then the only global symmetry which remains
in both theories is ${\bf Z}_{2N}\ $ spontaneously broken
down to
\ztwo by the respective bifermion condensates. This explains the existence
of $N$ vacua in both cases. We will compare the bosonic sectors of the
parent and daughter theories. Note that in the daughter theory
the part of the bosonic sector probed by the operators
of the type (\ref{veccur}), which have no analogs in the
parent theory, is inaccessible.

\begin{figure}[H]
  \begin{center}
\mbox{\kern-0.5cm
\epsfig{file=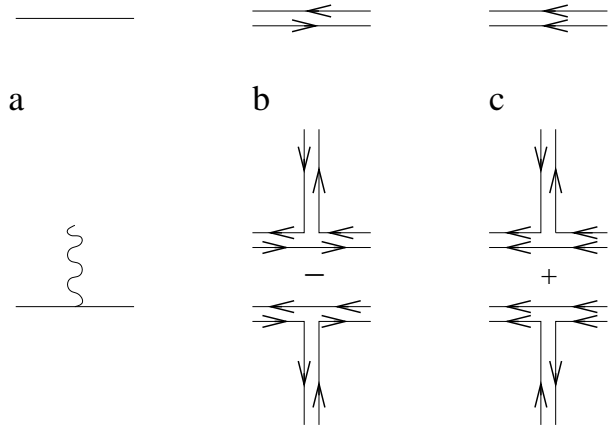,width=8.0true cm,angle=0}}
  \end{center}
\caption{(a) The fermionic propagator and the vertex. (b) \None
SYM. (c) The non-SUSY theory.}
\label{fey1}
\end{figure}

Let us start from  perturbative consideration. The general argument for
{\em any} orbifold/orientifold field theory is given in
\cite{Bershadsky:1998mb,Bershadsky:1998cb}.
Let us see how it works in our orientifold field theory.
The Feynman rules of the planar theory are shown in Fig.\ref{fey1}.
The difference between the orientifold theory and \None is that the
arrows on the fermionic lines point in the same direction, since
the fermion is in the antisymmetric representation, in contrast
to the supersymmetric theory where the gaugino is in the adjoint
representation and the arrows point in the opposite directions.
This difference between the two theories does not affect planar
graphs provided each
gaugino line is replaced by the sum
of $\lambda_{[..]}$ and $\lambda^{[..]}$.

\begin{figure}[H]
  \begin{center}
\mbox{\kern-0.5cm
\epsfig{file=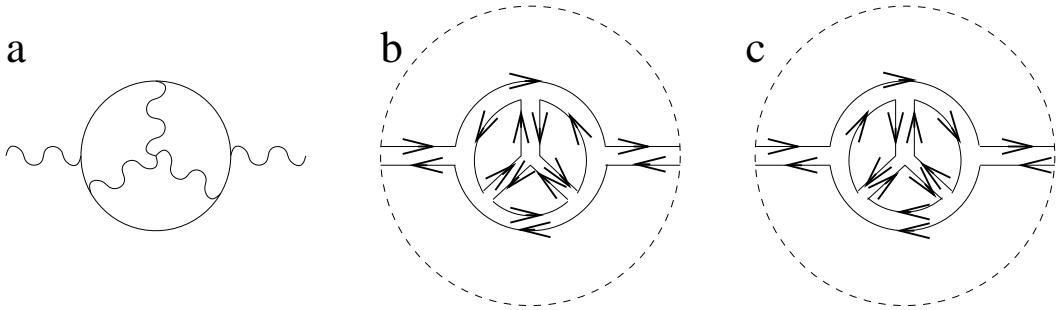,width=14.0true cm,angle=0}}
  \end{center}
\caption{(a) A typical planar contribution to the vacuum polarization.
(b) For \None SYM. (c) For the non-SUSY theory.}
\label{fey2}
\end{figure}

 There is a one-to-one correspondence between the planar graphs
of the two theories. Diagrammatically this works as follows (see, for
example, Fig. \ref{fey2}). Consider any planar diagram of the daughter theory:
by definition of planarity it can be drawn on a sphere. The fermionic
 propagators form closed, non-intersecting loops that divide the sphere into
regions. Each time we cross a fermionic line the orientation of color-indices
 loops (each one producing a factor $N$)
changes from clock to anti-clock wise, and vice-versa, as easily seen in
Fig. 2c. Thus, the fermionic loops
allow to attribute to each one of the above regions a binary label (say $\pm1$)
 according to whether the color loops go
clock- or anti-clock-wise in that region. Now
imagine that  the orientation
of color loops in all  regions with the $-1$ label is {\em reversed}. 
We will get a planar diagram of the SYM theory in which
all color loops go, by convention, clock-wise.
The number associated with both diagrams will be the same since the diagrams
inside each region always contain
an even number of powers of $g$ so that the relative minus signs of
Fig. 1 do not matter.

 Let us illustrate how this works, say, for
the inside part of the graph in Fig. \ref{fey2}b. In the parent theory
we have the color factor Tr $(T^aT^bT^c)\, f^{abc}$ while in the daughter one
Tr $(\bar T^a \bar T^b \bar T^c)\, f^{abc}$ where $\bar T = -\tilde T$
and the tilde marks the transposed matrix\,\footnote{To make the following expressions
concise we use a shorthand,
$T^a=T^a _{\Yfund} $ and $\bar T^a=T^a _{\overline{\Yfund}} $.
}. Using the fact that
$$
[T^aT^b]=i f^{abc}T^c\quad\mbox{and}\quad [\bar T^a\bar T^b]=i f^{abc}\bar T^c
$$
we immediately come to the conclusion that the above two expressions
coincide.

Thus, all perturbative results that we know of in \None SYM
apply for the orientifold model as well. For example, the
$\beta$ function of the orientifold field theory is
\beq
\beta = -\frac{1}{2\pi}\,\frac{3N\alpha^2}{1- (N\alpha)/(2\pi)}
\left\{1+ O\left(\frac{1}{N}\right)
\right\},
\eeq
where $\alpha = {g^2 \over 4\pi}$. In the large $N$ limit it coincides with the \None SYM result
\cite{nsvzbeta}. Note that the corrections are $1/N$ rather than
$1/N^2$. For instance, the exact first coefficient
of the $\beta$ function is $-3N-4/3$ versus
$-3N$ in the parent theory. 

Now let us argue that the perturbative argument can be elevated
to nonperturbative level in the case at hand. 
A heuristic argument in favor
 of the nonperturbative equivalence is that the coincidence
of all  planar graphs 
of the two theories implies that the relevant Casimir operators of the two
representations are equivalent in the large $N$ limit. The partition
functions of the two theories depend  on the   Casimir operators
 and therefore must  
coincide as well.   

A more formal line of reasoning is as follows.
It is essential that the fermion fields enter
bilinearly in the action, and that for any {\em given}
gauge field configuration in the parent theory there
is exactly the same configuration in the daughter one.
(The latter feature is absent in the \zk orbifolds.)
Our idea is to integrate out fermion fields
for any fixed gluon field configuration,
which yields respective determinants, and then compare them.

Consider the partition function of \None SYM,
\beq
{\cal Z}_0 = \int {\cal D}A\, {\cal D}\lambda \, \exp \left(iS[A,\lambda, J]
\right)\,,
\eeq
where $J$ is any source coupled to color-singlet  gluon operators
(we will discuss color-singlet fermion bilinears later).

For any given gluon field, upon
 integrating out the gaugino field, we obtain
\beq
{\cal Z}_0 = \int {\cal D}A \, \exp\left( iS[A, J]
\right)  \det \left (\slsh \partial + \slsh  A ^a T^a _{\rm adj}  \right )
\label{Z},
\eeq
where $T^a _{\rm adj}$ is a generator of the adjoint representation.

If one integrates out the fermion fields of the
non-supersymmetric orientifold theory, at fixed $A$,
 one arrives at a similar expression, but with the generators of the
anti-symmetric representation instead of the adjoint,
$T^a _{\rm adj} \rightarrow T^a _{\rm as}$. 

To compare the fermion determinants in the
parent and daughter theories (assuming that the gauge
field configuration $A_\mu^a (x),\,\,\, a=1,..., N^2,$ is the same and fixed)
we   must cast both fermion operators in  similar forms.
To this end we will   extend both theories.
In the parent one we introduce the second adjoint Weyl fermion
$\xi^i_j$ and combine two Weyl fermions into
one Dirac adjoint fermion $\Psi^i_j$. The determinant
in this extended theory is the square of the original one;
we assume that taking the square root at the end is harmless.

In the daughter theory instead of $\Psi^{[ij]}$, we will work
with the reducible representation, combining both symmetric and antisymmetric,
$\Psi^{ij}$
(no (anti)sym\-metrization over the color indices).
Then the number of the fermion Dirac fields is
the same as in the extended parent. Again, as in the previous case,
the determinant in the extended daughter theory is the square
of the original one.  For infrared regularization we
will introduce small mass terms,
assuming that the vanishing mass limit is smooth.

Next, we will make use of the fact that
\beq
T^a _{adj} \sim T^a _{\Yfund \, \times \, \overline{\Yfund}} =
T^a _{\Yfund} \otimes
1 + 1 \otimes T^a _{ \overline { \Yfund}}\,,
\label{mnone}
\eeq
and
\beq
T^a_{d}  = T^a _{\Yfund} \otimes
1 + 1 \otimes T^a_{\Yfund}\,,
\label{mn-two}
\eeq
or
\beq
T^a_{d}  = T^a_{\overline{\Yfund}} \otimes
1 + 1 \otimes T^a_{\overline{\Yfund}}\,.
\label{mntwo}
\eeq
The subscript d stands for ``daughter.''
Let us now introduce, as an auxiliary object,
\beq
F(A,B) = {\rm tr}\, \ln \left ( \slsh \partial + \slsh A^a ( T^a _{\Yfund} \otimes 1) 
 + \slsh B^a ( 1 \otimes T^a _{\overline{\Yfund}}) \right ) \,,
\label{mnthree} 
\eeq
and notice that (at large$N$)
$F$ is invariant under separate gauge
transformations acting on  $A^a$ and $B^a$. Using the fact that the
Wilson loop operators ${\cal W}_{\cal C}(A) = {\rm tr}\, P\, \exp
\left(
 i \int_{\cal C} A
\right)$
form
a
complete
set of gauge invariant operators, we 
can write 
\beq
\exp\left\{
F(A,B)\right\}
= \sum_{\cal
C,C'} \alpha _{\cal C,C'}{\cal W} _{\cal C} (A) {\cal W} _{\cal C'} (B).
\eeq
The partition function of \None SYM can be written, at large $N$, as
\beq
{\cal Z}_0 = \int {\cal D}A \,  {\cal D}B \, \delta (A-B)\, \exp
\left\{
-\smallfrac{1}{2}(S[A]+S[B]) \right\}
\sum _ {\cal
C,C'} \alpha _{\cal C,C'}{\cal W}_{\cal C} (A) {\cal W}_{\cal C'} (B).
\label{Z2}
\eeq
 Now let us turn
to the daughter theory. The partition function of the orientifold
theory, at large $N$, can be written as \eqref{Z2}, but with the orientations
of the Wilson loops ${\cal W} (B)$ {\em reversed},   since we replace
the fundamental fermions by anti-fundamental fermions. 
This gives
$$
{\cal W}_{\cal C'}\to {\cal W}_{\bar{\cal C'}} =
\left\{{\cal W}_{\cal C'}
\right\}^*\,.
$$
Moreover, ${\cal W}_{\cal C'}$ is real,
 and this is
why reversing the orientation should not change the value of the
partition function.

Let us present now a more detailed derivation
which relies on the fact that at large $N$
the two kinds of gluons do not interact with each other. The partition function
\eqref{Z2} can be written as
\beq
{\cal Z}_0 = \int {\cal D}A\, \exp \left\{-S[A] + F(A,B=A)\right\}.
\eeq
 Let us expand $\exp (-S)$
\beq
\exp  \left\{ -S[A]\right\}  = \sum _n {(-1)^n \over n!}  \left\{ S[A]\right\}^n .
\eeq
At $N \rightarrow \infty$, the  vertices or propagators coming from each
factor of $S$ cannot connect $A$ with $B$. Thus, we have
\beq
S^n \sim \sum _{k=0} ^n C_k S^k[A] S^{n-k}[B],
\eeq
where the $\sim$ sign means ``can be replaced by'' and $C_k$ 
is a combinatorial 
factor corresponding to the various 
choices of picking $k$ out of $n$, 
$$C_k = {n! \over k!(n-k)!}\,.
$$
Then
\begin{eqnarray}
\exp (-S) &\sim & \sum _n {(-1)^n \over n!} \sum _k {n! \over k!(n-k)!} 
\left\{ S[A]\right\} ^k
\left\{ S[B]\right\}^{n-k} 
\nonumber\\[3mm]
&=&
 \exp (- S[A]-S[B]). 
\label{action}
\end{eqnarray}
Thus, we conclude that $\exp (-S)$ acts as $\exp (-S[A]-S[B])$
and, at large $N$, we can think of the partition function
as if we actually have two gauge fields. Thus, at large $N$, Eq.~\eqref{Z2}
becomes 
\beq
{\cal Z}_0 = \sum _{\cal C,\cal C'} \alpha _{\cal C,\cal C'} \left\langle {\cal
W}_{\cal C}\right\rangle  \left\langle {\cal W}_{\cal C'}\right\rangle .
\label{factorization}
\eeq

 In order to pass to the
daughter theory one has to replace $B = B^a T^a _{\overline{\Yfund}}$
by $B = B^a T^a _{\Yfund}$. But $B$ is a dummy variable and, hence, this 
substitution will not change the value of the partition function.
Though we used a perturbative intuition in order to arrive to the
factorization of either the Wilson loops in \eqref{Z2} or the
action in Eq.~\eqref{action} we believe that this is an exact property of
the partition function. Note also that the factorization
in the partition function  
follows immediately from the existence of a master field \cite{witten}.
Thus, given the existence of  a large $N$ master field, the equivalence of
the parent and daughter theories is proven.

Another line of reasoning supporting non-perturbative planar equivalence
of the parent and orientifold theories
is based on lattice formulations of both theories.
If one examines strong coupling diagrams
for, say, a Wilson loop on the lattice, one readily concludes that the
large $N$
diagrams of the two theories can be put in a one-to-one correspondence
and agree to leading order in $1/N$.

For all these reasons we believe
that  the full equivalence of these two
theories takes place in the large $N$ limit, for  quantities which do not
involve fermion  external legs. The equivalence is
nonperturbative. An immediate result is that the vacuum energy
 of the orientifold theory is zero in the large $N$ limit (or at
finite $N$ it is $1/N$
relatively to a natural $N^2$ behavior). To be more precise, the expected
dependence of the vacuum energy on the UV cut-off and $N$ is generically
\beq
\rho =\Lambda _{UV} ^4 N^2 (c_0 + c_1/N + c_2/N^2 + ...).
\eeq
In the present theory $c_0=0$ while, generically, $c_1,c_2, ...$ are non-zero.
Namely, the vacuum energy is zero only in the planar theory. If one
keeps the combination $\Lambda _{UV} ^4 N^2$ fixed then the limiting
theory has a zero vacuum energy. While this is an almost trivial statement in the UV, simply
because of the equivalence of the planar graphs, it is highly nontrivial
from the IR point of view. The daughter theory hadronic spectrum consists
of bosons only, since it is impossible to form light color-singlet fermions.
Nonetheless one should remark that contributions to the cosmological constant
from color-singlet loops only enter at the level of the genus-1 (torus) diagrams.
These are already O($1/N^2$) down with respect to the leading contribution to
the cosmological constant (which vanishes) and O($1/N$) relative to the
presumed  leading
non-vanishing contribution in the daughter theory.

In order to go beyond the equivalence of vacuum diagrams
and correlators with external gluonic sources we have to understand
how the above argument
can be extended if we add fermionic bilinear sources.

It looks quite obvious that sources coupled to $\lambda\lambda \pm
\bar{\lambda} \bar{\lambda}$ in the parent theory
 can be mapped into sources coupled to
$\bar{\Psi}(1 \pm \gamma_5) \Psi$ in the daughter theory.
This will be enough to prove the equality of condensates in the two theories
and the $x$-independence of certain correlators, hence the parity-doublet
structure of the spectrum in both theories (see below). In particular, a
mass terms can be added to both theories without spoiling their large $N$
equivalence.
Although we have not made a systematic study of this problem it looks that
many other fermionic bilinears (FB)   --- involving the gluon field as 
well --- can be mapped
in the two theories so that
\beq
W_{\rm SYM} (J_{\rm glue}^{i}, J_{FB}^{j}) =
W_{\rm ori} (J_{\rm glue}^{i}, \tilde{J}_{FB}^{j}) (1 + O(1/N))
\eeq
with an explicit dictionary relating $J_{FB}^{j}$ and $\tilde{J}_{FB}^{j}$.
An example of such a pair is
\beq
\left( G_{\alpha\beta}\right)^i_j
 \left( \lambda_{\alpha}\right)^k_i \left( \lambda_{\beta}\right)^j_k
  \leftrightarrow \left( G_{\alpha\beta}\right)^i_j
\left(\lambda_{\alpha}\right)_{[ik]}\left(\xi_\beta \right)^{[kj]}\,,
\eeq
where $\left( G_{\alpha\beta}\right)^i_j$ is the gluon field strength tensor.

As was mentioned, we can derive the bifermion condensate
$\langle\lambda\xi\rangle \neq 0 $ in the daughter theory starting
from $\langle\lambda\lambda\rangle \neq 0 $ in the parent one, hence
$$
\langle\lambda\xi\rangle _k =N \Lambda ^3 \exp\left( i {2\pi k\over
N}\right)\,,
$$
 as in Eq. (\ref{condensate}).

Another property which is inherited by the daughter theory is
the constancy of the chiral correlator
\beq
\langle \lambda \xi (x_1)\,\,  \lambda \xi (x_2)\,\, ...\,\,
 \lambda \xi (x_n)\rangle =
{\rm const.}
\eeq
Physically it is related to the mass degeneracy of scalar and pseudoscalar
mesons.

It should be stressed that in \zk orbifolds
some background gauge field configurations are present in the parent
theory and absent in the daughter one,
so that comparison of the fermion determinants does
not  prove nonperturbative equivalence.  

To summarize,
we can relate many correlators which involve {\em even} number
of fermions and/or gluon operators (hence, the corresponding hadron spectra).
 However, the
parent and daughter theories differ from each other in the
sector of odd number of fermions. For example, there is a gauge
invariant three-fermions state in \None SYM. Such a state does not exist
 in the orientifold theory. Remember also that the sector of the
daughter theory probed by the operators of the type
(\ref{veccur}) is not accessible for predictions.

\section{The relation with type 0A string theory and M-theory}

\noindent

In this section we show that our orientifold field theory has a
simple realization in type 0A string theory.
Type 0A/B are bosonic closed string theories with a low-energy
spectrum which consist of the universal NS-NS sector, a tachyon and a
doubled set of R-R forms with respect to type IIA/B string theories.
Type 0B string theory has three kinds of orientifolds
\cite{Sagnotti:1995ga,Sagnotti:1996qj}. We will be mostly interested in the
nontachyonic one \cite{Sagnotti:1995ga,Sagnotti:1996qj}.
 Consider the action $\Omega
(-1)^{f_R}$, namely the world-sheet parity combined with
the world-sheet fermion
number. The NS-NS vacuum is odd with respect to this action and,
therefore, it is removed from the spectrum. Thus, the bulk theory is
tachyon free. In addition this orientifold removes half of the R-R
fields so that the theory now has only one set of R-R fields,
as in the type IIB theory. In order to remove R-R tadpoles
one has to introduce 32 D9 branes. The field theory on the D9 branes
is a 10 dimensional $U(32)$ gauge theory with an antisymmetric
fermion.

\begin{figure}
  \begin{center}
\mbox{\kern-0.5cm
\epsfig{file=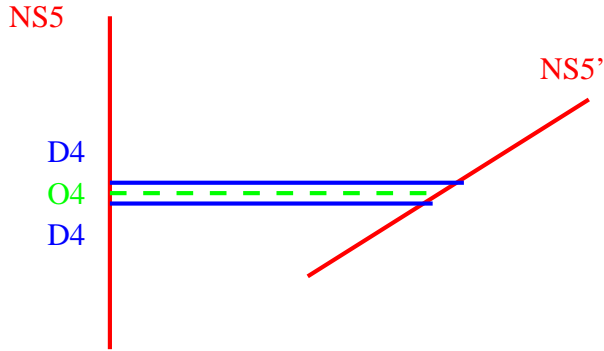,width=8.0true cm,angle=0}}
\label{N1}
  \end{center}
\caption{The type 0A brane configuration. The solid lines denote D4
branes. The dashed line denotes the orientifold 4-plane.}
\end{figure}

One can perform a sequence of T dualities to obtain a system of D4
branes and O4 plane. Moreover, a brane configuration which
consists of NS5 branes and an O4 plane as in Fig. \ref{N1},
leads \cite{Angelantonj:1999qg} to our ``orientifold field theory''.

We would like to use the relation of type 0A string theory to M-theory
to study the strong coupling regime of the orientifold
theory at large $N$. This argument is certainly not a proof and we
present it just to indicate that our results in Sect. 2
 are in agreement with
 the string theory picture.

Similarly to type IIA string theory, type 0A can be obtained from
M-theory by a compactification of the eleventh dimension on a
Scherk-Schwarz circle \cite{Bergman:1999km}. By using this conjecture
 we can lift the brane configuration (Fig. \ref{N1}) to M-theory,
similarly to the lift of the analogous type IIA brane configuration
\cite{Hori:1997ab,Witten:1997ep,Brandhuber:1997iy}.
Note that the presence of the
orientifold plane can be neglected in the large $N$
limit, since its R-R charge is negligible with respect to the R-R charge
of the branes ($N$). Similarly to the type IIA situation we will obtain
a smooth M5 brane and the resulting curve (the shape of the M5)
will be the same as the curve of \None SYM \cite{Brandhuber:1997iy}
\beq
S^N =1 .
\eeq
The meaning of this curve is that there are $N$ vacua with the
bifermion condensate \eqref{condensate} as the order parameter,
in agreement with our field
theory results.

\section{Discussion and conclusions}

\noindent

One of the goals  of this section is to compare in more detail
the results that are
obtained in this work for orientifold field theories with
the previous
results for orbifold field theories.

In order to be concrete let us discuss, as an example, the \ztwo
orbifold theory (see Table).
This theory as well has a realization in type 0A theory
\cite{Armoni:1999gc}. It lives on a brane configuration of type 0A
which consists of `electric' and `magnetic' D-branes,
hence the labels `e' and `m' in the Table.

Let us divide the operators in the \ztwo orbifold theory to operators
that are invariant (even) under the exchange of the labels `e' and
`m'
and  operators which are odd under this exchange. The first
are called ``untwisted operators,''  the second ``twisted operators.''
For example, the operator ${\cal O}^+={\rm tr}\, F_{\rm ee} ^2 + {\rm
tr}\, F_{\rm mm} ^2$ is an untwisted operator whereas
 ${\cal O}^-={\rm tr}\, F_{\rm ee} ^2 - {\rm
tr}\, F_{\rm mm} ^2$ is a twisted operator.

\begin{table}[H]
\begin{displaymath}
\begin{array}{l c@{ } c@{ } }
& \multicolumn{1}{c@{\times}}{U_e(N)}
& \multicolumn{1}{c@{}}{U_m(N)} \\
\hline
A _{\rm ee} & adj. & 1  \\
A _{\rm mm} & 1 & adj.  \\
\lambda _{\rm em} &  \Yfund & \overline{\Yfund}  \\
\lambda _{\rm me} &   \overline{\Yfund} & \Yfund  \\
\end{array}
\end{displaymath}
\caption{The \ztwo orbifold theory.}
\label{table_N=1}
\end{table}

The perturbative relation between the orbifold theory and its
supersymmetric parent concerns only the untwisted sector
\cite{Bershadsky:1998mb,Bershadsky:1998cb}. The parent theory does not
carry information about the twisted sector of the daughter theory.
It is always assumed that the vacuum of the daughter
theory is \ztwo invariant. However, the
\ztwo symmetry might be broken
dynamically\,\footnote{A.A. thanks Y. Shadmi for suggesting this
scenario.}, due to a an expectation value of ${\cal O}^-$.

A possible sign of the \ztwo instability
comes from perturbation theory. Indeed, let us assume
for a moment that at some UV scale where perturbation theory is
applicable the gauge couplings of the two U(N) factors in the
orbifold theory are slightly different. We will denote
$2\pi/\alpha_e =z_e$ and $2\pi/\alpha_m =z_m$
where the subscripts $e,\,\,m$ refer to the first and second U(N) factors,
respectively. It is not difficult to find the
renormalization group flow of $\delta z$ towards the IR domain.
As long as $\delta \alpha\ll \alpha_{e,m}$
we have
\beq
\frac{d (\delta z)}{d\,\ln \mu}= -\frac{3N^2}{z^2}\, (\delta z)
+\mbox{higher orders}\,.
\eeq
Neglecting weak logarithmic $\mu $ dependence of $\alpha$
we get
\beq
\delta z (\mu ) = \delta z (\mu_0 )\, \left(\frac{\mu}{\mu_0}
\right)^{-3N^2\alpha^2/(4\pi^2)}\,.
\eeq
If $\delta z$ is small in UV, it grows towards the IR domain,
an indication on a destabilization tendency. A similar analysis
for the conformal daughter theory of 
extended SUSY, \Nfour, was carried out  in Ref.~\cite{Klebanov:1999um}.

The advantage of the orientifold theory over the
orbifold theory is the absence of the twisted sector.
Moreover,
the gauge groups are the same in the
parent and daughter theories, and so are the patterns of the
spontaneous breaking
of the global
symmetry
and the numbers of vacua.
This is also seen from the string theory
 standpoint. The orbifold field theory originates from the type 0A
string theory which is obtained by a \ztwo orbifold of type IIA. The
result is a bosonic string theory with a tachyon in the twisted
sector. In contrast, the orientifold field theory originates from
a configuration with an orientifold which removes the twisted sector
(the result is very similar to the bosonic part of the type I string).

If the equivalence between \None SUSY gluodynamics and orientifold theory
 does hold nonperturbatively,
this must have a strong consequence for the
symmetry of the IR theory.
Indeed, the degeneracy of the meson masses
inherited from the parent theory would imply
that
in the daughter theory there is a tensorial  operator
(other than the energy-momentum tensor)
which is conserved in the large $N$ limit.
This is in no contradiction with the
Coleman-Mandula
 theorem  
\cite{Coleman:ad}
since in the large  $N$ limit
all scattering amplitudes vanish,
and the $S$ matrix tends to unity. Note that we discuss here only
 the large $N$ theory, namely the theory of planar graphs. In this
 limit the theory is expected to be a free theory of color singlet
glueballs. Glueball couplings are all suppressed by powers of $1/N$.

Since the perturbative planar equivalence
does not depend on the geometry of space-time,
one can compactify one or more dimensions,
with a small compactification size, to make the theory
weakly coupled. One can then compare the parent and the
orbifold theories.
This was  done, in
particular,  in Ref. \cite{Gorsky:2002wt}
where it was shown that the toron contributions
in the parent and \ztwo orbifold theories do not match
($T^4$ compactification is implied).

Compactifying one or more extra dimensions one should be very cautious,
however. In doing so we get a theory with scalar moduli
(flat directions at the classical level).
It may happen (and in fact happens)
that one or several components of the moduli fields
develop vacuum expectation values which scale as $N^{1/2}$.
This breaks (a part of) the gauge symmetry leading
to the Higgs regime. Simultaneously, the $1/N$ expansion is
broken too. Indeed, in this expansion we assume that
the smallness of $g^2$ is compensated
not by a large value of the gluon field
(the gluon propagator is $O(N^0)$) but by a large number of components
of the gluon fields circulating in loops.
Upon compactification
we get scalar fields, just a few components of which may condense,
compensating the smallness of $g^2$.

An example was given
by Tong \cite{Tong:2002vp}, who considered $R^3\times S$
compactification at one loop. On $R^3\times S$
the third spatial component
 of the gluon field becomes
a scalar field with a flat (vanishing)
potential at the classical level, $(A_3)^i_j\to \phi^i_j$.
(Alternatively, one may speak of
the Polyakov line in the $x_3$ direction.)  In the parent theory
the flatness is (perturbatively) maintained to all orders by supersymmetry.
Nonperturbatively, the flatness is lifted
by $R^3\times S$ instantons (monopoles)
which generate a superpotential. It turns
out that in supersymmetric vacua
the non-Abelian
gauge symmetry is completely broken
by $\langle\phi^i_j\rangle = v_i\delta^i_j$, down
to U(1)$^{2N}$, so that the theory is in the
Coulomb phase \cite{Davies:1999uw}.
The expectation values
$v_i$ scale as $1/g\sim \sqrt N$.
This explains why the generated masses $\sim g v_i\sim N^0$.

Tong showed
that in the daughter theory
a potential emerges at one loop, making
 the point
of the broken gauge symmetry  unstable.
Shifting from this point, for a trial,
one finds oneself in the \ztwo -noninvariant
(or twisted) sector, for which no planar equivalence exists,
and which proves to be energetically favored in this case.
In the true vacuum the energy density is
negative rather than zero and the full gauge symmetry
is restored. The daughter theory
is in the  confining phase. Obviously, then there is no equivalence.

The above remark implies that
in considering equivalence
between ${\cal N}=2$ (or ${\cal N}=4$) theories
where scalar moduli are abundant, with the corresponding
orbifold/orientifold theories,
one should be sure to be in the  confining rather than Higgs regime.

Another indication that the presence of the twisted sector
in \zk orbifolds may have a negative impact on the untwisted sector
came from consideration of the low-energy theorems.
In particular, topological
susceptibilities in the parent and \ztwo daughter theories
were analyzed in \cite{Gorsky:2002wt}.

The topological susceptibility reflects dependence of the vacuum
 energy on the
vacuum angle $\theta$. For massless fermions such dependence is absent
and the topological susceptibility vanishes.
Only if $m\neq 0$, the topological susceptibility does not
vanish and can be readily derived to
leading order in $m$.
Thus it is necessary to deform the parent/daughter theories by
fermion mass terms. This deformation does not affect perturbative planar
equivalence.

It was shown \cite{Gorsky:2002wt} that under certain reasonable
assumptions the  topological susceptibilities
do not match, the discrepancy being a factor of 2.
This factor can be traced back to the fact that the number of vacua
in U($2N$) SUSY parent is $2N$
while its  $U_{\rm e}(N)\times U_{\rm m}(N)$
daughter (\ztwo orbifold) has only $N$ vacua.

Needless to say,  in our case of the orientifold daughter
the topological susceptibilities are identical,
as so are the gauge groups, gauge couplings and the number
of vacua.

In conclusion, let us formulate a question
which naturally comes to one's mind at the end of this paper:

``What is the symmetry of the daughter theory,  weaker than SUSY,
which nevertheless implies infinite number of
degeneracies in the spectrum?''

\Acknowledgements

A.A. would like to thank O. Aharony, C. Angelantonj,
O. Bergman, Y. Frishman, A. Hanany, R. Rabadan, E. Rabinovici,
 and Y. Shadmi
for useful discussions.
M.S. is grateful to A. Gorsky and D. Tong
for numerous communications and discussions of the
\zk orbifold theories. The work of M.S. is supported in
part  by DOE grant DE-FG02-94ER408.


\begin{thebibliography}{99}

\bibitem{srev}
N.~Seiberg,
Proceedings 4th Int. Symposium on  Particles, Strings, and Cosmology
(PASCOS 94),  Syracuse, New York,  May 1994, Ed.
 K.  C. Wali (World
   Scientific,  Singapore, 1995), p. 183
[hep-th/9408013];
Int.\ J.\ Mod.\ Phys.\ A {\bf 12} (1997) 5171
[hep-th/9506077].

\bibitem{Strassler:2001fs}
M.~J.~Strassler,
{\em On methods for extracting exact
 nonperturbative results in  non-supersymmetric gauge theories,}
hep-th/0104032.


\bibitem{Gorsky:2002wt}
A.~Gorsky and M.~Shifman,
Phys.\ Rev.\ D {\bf 67}, 022003 (2003)
[hep-th/0208073].

\bibitem{Tong:2002vp}
D.~Tong,
{\em Comments on condensates in non-supersymmetric orbifold field theories,}
hep-th/0212235.


\bibitem{Dijkgraaf:2002wr}
R.~Dijkgraaf, A.~Neitzke and C.~Vafa,
{\em Large N strong coupling dynamics in non-supersymmetric
orbifold field  theories,}
hep-th/0211194.

\bibitem{Veneziano:1982ah}
G.~Veneziano and S.~Yankielowicz,
Phys.\ Lett.\ B {\bf 113}, 231 (1982).

\bibitem{Bershadsky:1998mb}
M.~Bershadsky, Z.~Kakushadze and C.~Vafa,
Nucl.\ Phys.\ B {\bf 523}, 59 (1998)
[hep-th/9803076].

\bibitem{Bershadsky:1998cb}
M.~Bershadsky and A.~Johansen,
Nucl.\ Phys.\ B {\bf 536}, 141 (1998)
[hep-th/9803249].

\bibitem{Armoni:1999gc}
A.~Armoni and B.~Kol,
JHEP {\bf 9907}, 011 (1999)
[hep-th/9906081].

\bibitem{Angelantonj:1999qg}
C.~Angelantonj and A.~Armoni,
Nucl.\ Phys.\ B {\bf 578}, 239 (2000)
[hep-th/9912257].

\bibitem{nsvzbeta}
V.~A.~Novikov, M.~A.~Shifman, A.~I.~Vainshtein and V.~I.~Zakharov,
Nucl.\ Phys.\ B {\bf 229}, 381 (1983);
Phys.\ Lett.\ B {\bf 166}, 329 (1986).

\bibitem{witten}
E. Witten, in
{\em Recent Developments In Gauge Theories},
 Proceedings of the 1979 Cargese Summer
 Institute, Eds. G.~'t Hooft, {\em et al.} (Plenum Press, 1980).

\bibitem{Sagnotti:1995ga}
A.~Sagnotti,
{\em Some properties of open string theories,}
hep-th/9509080.

\bibitem{Sagnotti:1996qj}
A.~Sagnotti,
Nucl.\ Phys.\ Proc.\ Suppl.\  {\bf 56B}, 332 (1997)
[hep-th/9702093].

\bibitem{Bergman:1999km}
O.~Bergman and M.~R.~Gaberdiel,
JHEP {\bf 9907}, 022 (1999)
[hep-th/9906055].

\bibitem{Hori:1997ab}
K.~Hori, H.~Ooguri and Y.~Oz,
Adv.\ Theor.\ Math.\ Phys.\  {\bf 1}, 1 (1998)
[hep-th/9706082].

\bibitem{Witten:1997ep}
E.~Witten,
Nucl.\ Phys.\ B {\bf 507}, 658 (1997)
[hep-th/9706109].

\bibitem{Brandhuber:1997iy}
A.~Brandhuber, N.~Itzhaki, V.~Kaplunovsky, J.~Sonnenschein and S.~Yankielowicz,
Phys.\ Lett.\ B {\bf 410}, 27 (1997)
[hep-th/9706127].

\bibitem{Klebanov:1999um}
I.~R.~Klebanov,
Phys.\ Lett.\ B {\bf 466}, 166 (1999)
[hep-th/9906220].

\bibitem{Davies:1999uw}
N.~M.~Davies, T.~J.~Hollowood, V.~V.~Khoze and M.~P.~Mattis,
Nucl.\ Phys.\ B {\bf 559}, 123 (1999)
[hep-th/9905015].

\bibitem{Coleman:ad}
S.~R.~Coleman and J.~Mandula,
Phys.\ Rev.\  {\bf 159}, 1251 (1967).

\end{thebibliography}
\end{document}